\documentclass[journal,9pt]{IEEEtran}

\usepackage{amsmath,amsfonts}
\usepackage{algorithmic}
\usepackage{algorithm}
\usepackage{array}

\usepackage[caption=false,font=normalsize,labelfont=sf,textfont=sf]{subfig}

\usepackage{textcomp}
\usepackage{stfloats}
\usepackage{url}
\usepackage{verbatim}
\usepackage{graphicx}
\usepackage{cite}
\usepackage{hyperref}
\begin{document}

\title{Virtual VNA 2.0: Ambiguity-Free Scattering Matrix Estimation by Terminating Not-Directly-Accessible Ports \\with Tunable and Coupled Loads}
%
%
%

\author{Philipp~del~Hougne,~\IEEEmembership{Member,~IEEE}
\thanks{The author acknowledges funding from the IETR PEPS program (project ``IMPEST''), the ANR France 2030 program (project ANR-22-PEFT-0005), the European Union's European Regional Development Fund, and the French region of Brittany and Rennes Métropole through the contrats de plan État-Région program (projects ``SOPHIE/STIC \& Ondes'' and ``CyMoCoD'').}
\thanks{P.~del~Hougne is with Univ Rennes, CNRS, IETR - UMR 6164, F-35000 Rennes, France (e-mail: philipp.del-hougne@univ-rennes.fr).}
}

\maketitle

\begin{abstract}
We recently introduced the ``Virtual VNA'' concept which estimates the $N \times N$ scattering matrix characterizing an \textit{arbitrarily complex} linear reciprocal system with $N$ monomodal lumped ports by inputting and outputting waves only via $N_\mathrm{A}<N$ ports while terminating the $N_\mathrm{S}=N-N_\mathrm{A}$ remaining ports with known tunable individual loads. However, vexing ambiguities about the signs of the off-diagonal scattering coefficients involving the $N_\mathrm{S}$ not-directly-accessible (NDA) ports remained. If only phase-insensitive measurements were used, an additional blockwise phase ambiguity ensued.
Here, inspired by the emergence of ``beyond-diagonal reconfigurable intelligent surfaces'' in wireless communications, we lift all ambiguities with at most $N_\mathrm{S}$ additional measurements involving a known multi-port load network. We experimentally validate our approach based on an 8-port chaotic cavity, using a simple coaxial cable as two-port load network. Endowed with the multi-port load network, the ``Virtual VNA 2.0'' is now able to estimate the entire scattering matrix without any ambiguity, even without ever measuring phase information explicitly. 
Potential applications include the challenging characterization of large and/or embedded antenna arrays.
\end{abstract}


\section{Introduction}
\IEEEPARstart{A}{n} \textit{arbitrarily complex} linear passive time-invariant scattering system with $N$ monomodal lumped ports is fully characterized by its $N\times N$ scattering matrix. Traditionally, the latter is routinely and directly measured with an $N$-port vector network analyzer (VNA) for frequencies ranging from the radio-frequency via the microwave to the millimeter-wave regime. However, in important applications, it is desirable to be able to estimate the full scattering matrix without attaching a transmission line to inject and receive waves via each of the system's ports: (i) the number of the system's ports may drastically exceed the number of available VNA ports when characterizing large antenna arrays; (ii) some ports may be embedded (e.g., inside a circuit~\cite{zhang2019overview} or a biological system) and thus inaccessible; (iii) the attachment of a transmission line to the port may perturb the scattering properties of the system (e.g., a miniaturized antenna~\cite{icheln1999reducing,skrivervik2001pcs}). We hence partition the system's ports into accessible and not-directly-accessible (NDA) ports.
We define an NDA port as a port via which we do not inject/receive waves (due to any of the three aforementioned reasons); we further assume that we can terminate an NDA port with tunable loads.
Is it possible to conceive a ``virtual VNA'' that retrieves the system's full scattering matrix by measuring the ``measurable'' scattering matrix at the $N_\mathrm{A}$ accessible ports for different terminations of the $N_\mathrm{S}$ NDA ports?

Related questions have been studied in metrology for decades under terms like ``unterminating''~\cite{bauer1974embedding} (only studied for $N_\mathrm{S}=1$) and ``port reduction''~\cite{lu2000port,lu2003multiport,pfeiffer2005recursive}. Port-reduction methods chiefly differ from the problem we address in the present work in that ports change between being accessible or NDA over the course of the studied methods.\footnote{For the same reason, related recent efforts~\cite{buck2022measuring} based on embedded element radiation patterns are different from the problem we address here.} 
For $N_\mathrm{S}=1$, the literature contains multiple experimental works tackling  (sometimes in special scenarios) parts of the problem we address, in contexts spanning metrology and antenna characterization~\cite{garbacz1964determination,bauer1974embedding,mayhan1994technique,davidovitz1995reconstruction,pfeiffer2005recursive,pfeiffer2005equivalent,pfeiffer2005characterization,pursula2008backscattering,bories2010small,van2020verification,sahin2021noncontact,kruglov2023contactless}. 
Two works~\cite{wiesbeck1998wide,monsalve2013multiport} studied parts of the problem we address in special scenarios with $N_\mathrm{S}=2$. 

Recently, Refs.~\cite{del2024minimal,Shilinkov_Maaskant_2024} tackled the question of interest raised in the first paragraph in its entirety, considering the generic case of an \textit{arbitrarily complex} system in Ref.~\cite{del2024minimal}. It turned out that the full scattering matrix can be retrieved \textit{up to sign ambiguities on off-diagonal terms associated with NDA ports if the terminations of the NDA ports can be switched between at least three distinct known load impedances}. A single directly accessible port is sufficient~\cite{Shilinkov_Maaskant_2024} but at least two directly accessible ports reduce the upper bound on the number of required load configurations by $N_\mathrm{S}(N_\mathrm{S}-1)/2$~\cite{del2024minimal}. On the one hand, there is a closed-form approach, developed and numerically validated for the particular case of $N_\mathrm{A}=1$ in Ref.~\cite{Shilinkov_Maaskant_2024} and developed for the general case of $N_\mathrm{A}>1$ and experimentally validated in Ref.~\cite{del2024minimal}. The closed-form approach requires specific load configurations and thereby defines an upper bound on the number of required load configurations. Using more directly accessible ports improves the robustness to measurement noise. On the other hand, there is a gradient-descent approach compatible with (potentially opportunistic) random load configurations, developed and experimentally validated in Ref.~\cite{del2024minimal}. The gradient-descent approach is generally more robust to measurement noise because the number of loads whose configurations change between measurements can strongly exceed one or two (for large $N_\mathrm{S}$). Moreover, the gradient-descent approach can recover the full scattering matrix purely based on intensity-only data, except for a blockwise phase ambiguity in addition to the aforementioned sign ambiguities.  None of the approaches requires specific characteristics of the three loads nor that they are the same at each NDA port~\cite{del2024minimal}.

Both fundamentally and for technological applications, an important question hence remains: how can these vexing ambiguities be lifted? Direct transmission measurements from \textit{only one} accessible port to all NDA ports lift the ambiguities~\cite{del2024minimal} but are not allowed if the NDA ports are ``truly NDA''. In special cases, one can use a priori knowledge, e.g., about the system's characteristics near dc~\cite{lu2000port,lu2003multiport,pfeiffer2005recursive} or the system's geometric details~\cite{Shilinkov_Maaskant_2024}. Our goal, however, is a generically valid approach that can be applied to arbitrarily complex systems without any a priori knowledge. 

The pivotal ingredient for a generic ambiguity-free approach is a \textit{multi-port} load network (MPLN). All aforementioned approaches limit themselves to terminating NDA ports with individual one-port loads. The importance of using an MPLN was already recognized in Ref.~\cite{denicke2012application} although this work only solved parts of the problem of estimating the \textit{full} scattering matrix without \textit{any} ambiguities that we address. MPLNs also appear in other contexts such as a variation of the port-reduction method~\cite{lin2015multiport} in metrology and beyond-diagonal reconfigurable intelligent surfaces (BD-RIS)~\cite{shen2021modeling,BDRIS_renzo_clerckx_2024,del2024physics} in wireless communications.

The present work clarifies that ambiguity-free physics-compliant end-to-end channel estimation in RIS-parametrized radio environments requires specific BD-RIS architectures and is not possible with conventional RIS. However, eliminating  ambiguities is \textit{not} required in the RIS context~\cite{sol2023experimentally}, nor for optimal non-invasive focusing on a perturbation-inducing target inside a complex medium~\cite{sol2024optimal}.

Here, we demonstrate theoretically and experimentally that a proper solution to the ambiguity problem (fully in line with the NDA requirement) consists in using at least one known MPLN in addition to the three known tunable individual loads. The MPLN has at least two ports and can simply be a coaxial cable. In this paper, after performing one of the three procedures involving reconfigurable individual loads as outlined in Ref.~\cite{del2024minimal}, pairs of NDA ports are successively connected to a coaxial cable; in addition, once an NDA port and an accessible port are connected to a coaxial cable. The use of coupled loads lifts all ambiguities (even in the case of intensity-only measurements) and thereby crowns the ``virtual VNA'' concept which can now operate free of any ambiguity, and even without ever measuring phase information. 

\textit{Organization:} In Sec.~\ref{sec_overview}, we provide an overview of the ``Virtual VNA 2.0'' principle and its potential embodiment in two applications. 
In Sec.~\ref{sec_theory}, we propose a procedure based on a two-port load network (2PLN) to remove the sign ambiguity (Sec.~\ref{sec_sign_ambiguity}), as well as the additional blockwise phase ambiguity in the case of intensity-only measurements (Sec.~\ref{sec_phase_ambiguity_removal}). In Sec.~\ref{sec_exp}, we experimentally validate the proposed procedures. We close with a brief discussion (Sec.~\ref{sec_discussion}) and conclusion (Sec.~\ref{sec_conclusion}).

\begin{figure}[b!]
    \centering
    \includegraphics[width=\columnwidth]{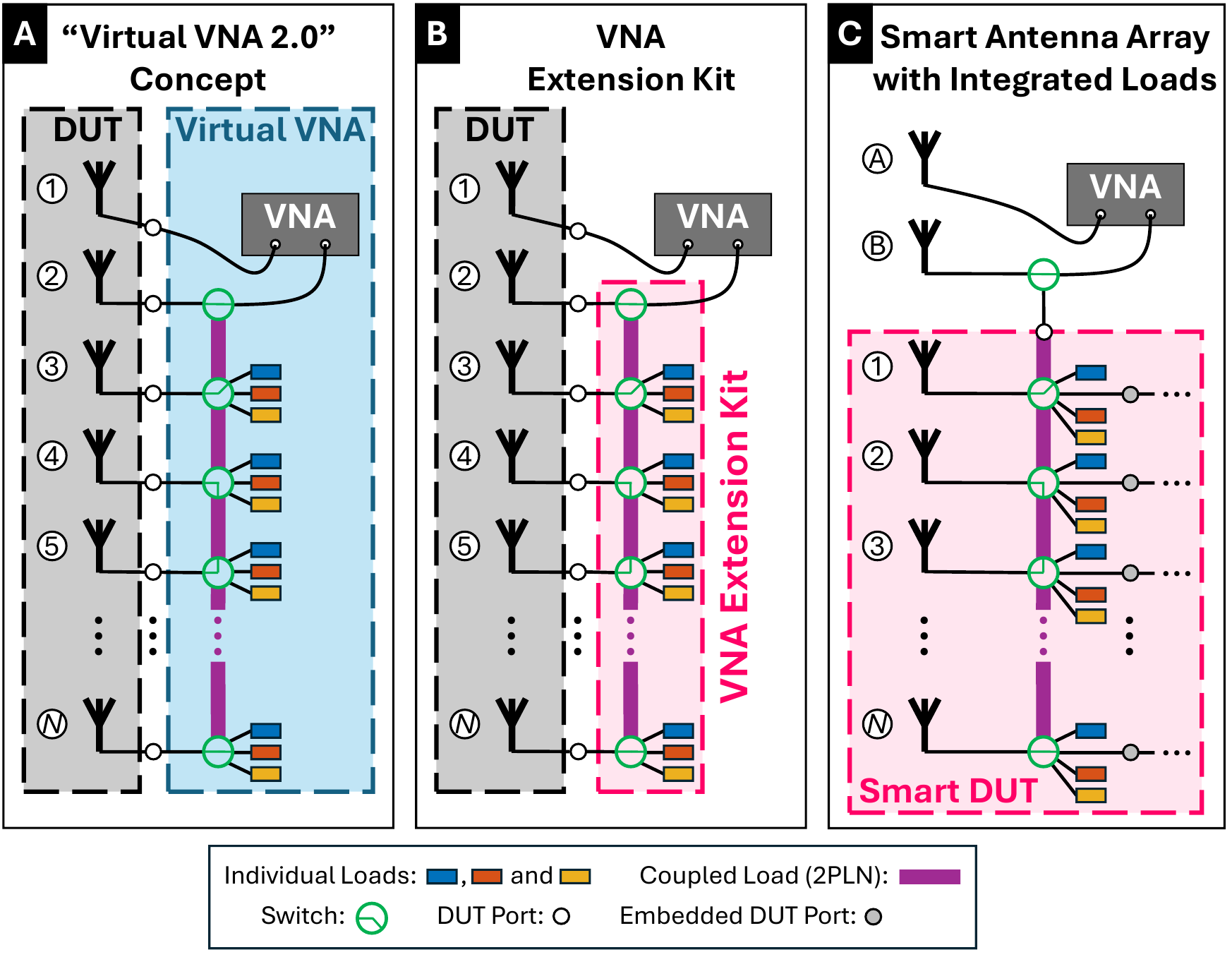}
    \caption{Schematic of the setup of the proposed ambiguity-free ``Virtual VNA 2.0'' for the characterization of an $N$-element antenna array using an $N_\mathrm{A}$-port VNA (for $N_\mathrm{A}=2$). (A) Concept. (B) Implementation via VNA extension kit. (C) Implementation via integration into antenna array.}
    \label{Fig_sketch}
\end{figure}

\section{Principle}
\label{sec_overview}

We consider a device under test (DUT) with $N$ lumped monomodal ports. 
Our only assumptions about the DUT are linearity, passivity, time-invariance and reciprocity; for a generalization to non-reciprocal systems, the reader is referred to~\cite{V2NA_3p0}. Most antenna arrays satisfy these requirements -- irrespective of the antenna design and the potential presence of scattering objects.
Our goal is to determine the DUT's $N\times N$ scattering (or impedance) matrix without connecting each of its ports directly to a VNA. Instead, we use an $N_\mathrm{A}$-port VNA, where $N_\mathrm{A}\ll N$, as illustrated in Fig.~\ref{Fig_sketch}A for $N_\mathrm{A}=2$.
The $N_\mathrm{A}$ ``accessible'' DUT ports are connected to the $N_\mathrm{A}$-port VNA; the remaining $N_\mathrm{S}=N-N_\mathrm{A}$ ``not-directly-accessible'' (NDA) DUT ports are connected to a reconfigurable load network. The latter enables us to terminate each of the $N_\mathrm{S}$ NDA ports with three distinct individual loads (e.g., in Fig.~\ref{Fig_sketch}A the switch associated with the third [$N$th] DUT port is configured to connect to the blue [red] load) or to connect any pair of neighboring DUT ports via a 2PLN (e.g., in Fig.~\ref{Fig_sketch}A the switches associated with the fourth and fifth DUT ports are configured to connect to a purple 2PLN). 
The 2PLN distinguishes the ``Virtual VNA 2.0'' concept from the ``Virtual VNA'' introduced in~\cite{del2024minimal}. 
In addition, one of the DUT's accessible ports can be connected to one of the DUT's NDA ports via a 2PLN (e.g., in Fig.~\ref{Fig_sketch}A the second and third DUT ports can be connected via a purple 2PLN).

Our general methodology consists in measuring the complex-valued scattering matrix at the DUT's accessible ports (using an $N_\mathrm{A}$-port VNA) for a series of configurations of terminations of the DUT's NDA ports via the tunable load network. The required set of configurations depends on the approach (closed form vs. gradient descent, see illustration in Fig.~\ref{Fig1}E) and is discussed later. 
We also consider an approach that only requires intensity-only measurements of waves exiting the accessible ports upon injecting different known coherent wavefronts into the accessible ports. In principle, the required setup would involve a set of coherent sources, circulators, and a multi-channel spectrum analyzer; however, we can emulate such measurements in software based on one measurement (with a VNA) of the scattering matrix at the accessible ports.

The ``Virtual VNA 2.0'' principle can find (at least) two distinct applications to antenna array characterization. The first application concerns the characterization of a large antenna array with many ports that are physically accessible, but their number largely exceeds the number of available VNA ports, such that the vast majority of antenna ports is effectively NDA for practical purposes (despite the availability of physical access). 
In that case, as illustrated in Fig.~\ref{Fig_sketch}B, one can imagine connecting the DUT's NDA ports to a VNA extension kit that implements the basic requirements of the ``Virtual VNA 2.0'' concept demonstrated in the present paper, as very recently reported in~\cite{KitVNA}. Importantly, the number of required switches scales linearly with $N_\mathrm{S}$, in contrast to existing commercially available full-crossbar switch matrix solutions. 
The second application concerns antenna arrays with embedded ports that are physically inaccessible, hence NDA. In that case, as illustrated in Fig.~\ref{Fig_sketch}C, one can imagine integrating the basic ingredients of the ``Virtual VNA 2.0'' concept directly into the antenna array and use auxiliary antennas with accessible ports (labeled A and B in Fig.~\ref{Fig_sketch}C). The integration of switches and individual tunable loads required in both applications is technically feasible, as already experimentally demonstrated in works such as Refs.~\cite{pfeiffer2005characterization,pfeiffer2005equivalent,kruglov2023contactless}. Meanwhile, the 2PLNs can be realized as simple waveguide connections between pairs of neighboring switches.

\section{Theory of Ambiguity Removal with Coupled Loads}
\label{sec_theory}

In this section, we describe the theoretical considerations for removing ambiguities with coupled loads. In Sec.~\ref{sec_sign_ambiguity}, we address the inevitable sign ambiguities arising in the case of complex-valued measurements with individual tunable loads. In Sec.~\ref{sec_phase_ambiguity_removal}, we address the additional inevitable blockwise phase ambiguity arising in the case of phase-insensitive measurements with individual tunable loads. 
The required load configurations are shown in Fig.~\ref{Fig1}E.

We denote by $\mathcal{A}$ and $\mathcal{S}$  the sets of accessible and NDA port indices, respectively. Subscripts $\mathcal{AA}$, $\mathcal{AS}$ and $\mathcal{SS}$ hence identify blocks of the DUT's scattering (or impedance) matrix.

\subsection{Sign Ambiguity Removal}
\label{sec_sign_ambiguity}

Considering the case of complex-valued measurements, and assuming that either the closed-form or gradient-descent-based method detailed in Ref.~\cite{del2024minimal} has been performed (without the ``optional'' sign ambiguity removal mentioned in Ref.~\cite{del2024minimal} based on directly accessing the NDA ports), in the present section we exclusively discuss how to remove the sign ambiguity of such an estimate using additional measurements involving a 2PLN. We detail the procedure in terms of impedance parameters but an equivalent implementation in terms of scattering parameters exists, as shown in Sec.~\ref{sec_phase_ambiguity_removal}.

\subsubsection{Background on sign ambiguity}
\label{sub_sec_background_sign_ambiguity}

To clearly understand the origin of the sign ambiguities and to appreciate the importance of the 2PLN in lifting them, let us consider a generic $M$-port system characterized by its impedance matrix $\check{\mathbf{Z}}\in\mathbb{C}^{M \times M}$. Two ports are terminated by a 2PLN characterized by its impedance matrix
\begin{equation}
\check{\mathbf{Z}}^\mathrm{2PLN} = \begin{bmatrix}    a & b \\ c &  d     \\ \end{bmatrix} \in\mathbb{C}^{2 \times 2},
\end{equation}
where $b=c$ if the 2PLN is reciprocal; $b=c=0$ if the 2PLN is composed of two individual (i.e., uncoupled) loads. The 2PLN terminates the $k$th and $l$th ports of the $M$-port system, and the set $\mathcal{F}$ contains the port indices of the remaining $M-2$ ports. Using the convenient partition
\begin{equation}
    \check{\mathbf{Z}} = \begin{bmatrix}  \check{\mathbf{Z}}_\mathcal{FF}  &  \check{\mathbf{z}}_k &  \check{\mathbf{z}}_l \\ \check{\mathbf{z}}_k^T & \check{\zeta}_k & \check{\kappa}_{kl}   \\ \check{\mathbf{z}}_l^T  & \check{\kappa}_{kl} & \check{\zeta}_l   \\ \end{bmatrix},
    \label{eq2}
\end{equation}
the measurable impedance matrix $\hat{\check{\mathbf{Z}}}\in\mathbb{C}^{(M-2) \times (M-2)}$ is~\cite{del2024minimal,prod2024efficient}
\begin{equation}
\begin{split}
    \hat{\check{\mathbf{Z}}} &= {\check{\mathbf{Z}}}_\mathcal{FF} - \begin{bmatrix}    \check{\mathbf{z}}_k &  \check{\mathbf{z}}_l \end{bmatrix} \left( \begin{bmatrix}    \check{\zeta}_k & \check{\kappa}_{kl} \\ \check{\kappa}_{kl} &  \check{\zeta}_l     \\ \end{bmatrix} + \begin{bmatrix}    a & b \\ c &  d     \\ \end{bmatrix} \right)^{-1} \begin{bmatrix}    \check{\mathbf{z}}_k^T \\  \check{\mathbf{z}}_l^T \\ \end{bmatrix} \\ &= {\check{\mathbf{Z}}}_\mathcal{FF} - \begin{bmatrix}    \check{\mathbf{z}}_k &  \check{\mathbf{z}}_l \end{bmatrix} \left( \check{\chi} \begin{bmatrix}    \check{\zeta}_l+d & -\left(\check{\kappa}_{kl}+b\right) \\ -\left(\check{\kappa}_{kl}+c\right) &  \check{\zeta}_k+a     \\ \end{bmatrix}  \right) \begin{bmatrix}    \check{\mathbf{z}}_k^T \\  \check{\mathbf{z}}_l^T \\ \end{bmatrix},
    \end{split}
    \label{eq3}
\end{equation}
where 
\begin{equation}
    \check{\chi} = \left((\check{\zeta}_k+a)(\check{\zeta}_l+d) - (\check{\kappa}_{kl}+b)(\check{\kappa}_{kl}+c)\right)^{-1}.
\end{equation}
Assuming that the 2PLN is reciprocal for simplicity, we can rewrite Eq.~(\ref{eq3}) as follows:
\begin{equation}
\begin{aligned}
   \hat{\check{\mathbf{Z}}} = {\check{\mathbf{Z}}}_\mathcal{FF} -  \check{\chi} \left[ (\check{\zeta}_l+d)\check{\mathbf{z}}_k\check{\mathbf{z}}_k^T + (\check{\zeta}_k+a)\check{\mathbf{z}}_l\check{\mathbf{z}}_l^T \right. \\
   \left. - (\check{\kappa}_{kl}+b) \left(\check{\mathbf{z}}_k\check{\mathbf{z}}_l^T + \check{\mathbf{z}}_l\check{\mathbf{z}}_k^T \right) \right].
\end{aligned}
\label{eq5}
\end{equation}

All entries of the estimates of the vectors $\check{\mathbf{z}}_k$ and $\check{\mathbf{z}}_l$ have the same sign ambiguity $\pm_{\check{\beta}_k}$ and $\pm_{\check{\beta}_l}$, respectively~\cite{del2024minimal}. 
The terms in Eq.~(\ref{eq5}) involving $\check{\mathbf{z}}_k\check{\mathbf{z}}_k^T$ and $\check{\mathbf{z}}_l\check{\mathbf{z}}_l^T$ are inevitably insensitive to the sign of $\check{\mathbf{z}}_k$ (and thereby to $\pm_{\check{\beta}_k}$) and the sign of $\check{\mathbf{z}}_l$ (and thereby to $\pm_{\check{\beta}_l}$), respectively. 

If $b=0$, then the term $\check{\kappa}_{kl}\left(\check{\mathbf{z}}_k\check{\mathbf{z}}_l^T + \check{\mathbf{z}}_l\check{\mathbf{z}}_k^T \right)$ cannot help to lift the sign ambiguities $\pm_{\check{\beta}_k}$ and $\pm_{\check{\beta}_l}$ because $\check{\kappa}_{kl}$ is itself not known without ambiguity. Instead, 
\begin{equation}
\pm_{\check{\kappa}_{kl}} = \pm_{\check{\beta}_k}\,\pm_{\check{\beta}_l}.
\label{eq666}
\end{equation}
Moreover, $\check{\chi}$ is insensitive to the sign of $\check{\kappa}_{kl}$ (and thereby to $\pm_{\check{\kappa}_{kl}}$).

In contrast, if $b \neq 0$, the term $\check{\chi}$ is sensitive to $\pm_{\check{\kappa}_{kl}}$; moreover, the magnitude of the term $(\check{\kappa}_{kl}+b) \left(\check{\mathbf{z}}_k\check{\mathbf{z}}_l^T + \check{\mathbf{z}}_l\check{\mathbf{z}}_k^T \right)$ is sensitive to $\pm_{\check{\kappa}_{kl}}$. Hence, $b \neq 0$ allows us to determine $\pm_{\check{\kappa}_{kl}}$ by predicting $\hat{\check{\mathbf{Z}}}$ for the two possible values of $\pm_{\check{\kappa}_{kl}}$ and retaining the value whose corresponding estimate of  $\hat{\check{\mathbf{Z}}}$ is closer to the corresponding measurement thereof. Having determined $\pm_{\check{\kappa}_{kl}}$ imposes constraints on the possible combinations of values that $\pm_{\check{\beta}_k}$ and $\pm_{\check{\beta}_l}$ can take, as seen in Eq.~\ref{eq666}, but is not enough to determine $\pm_{\check{\beta}_k}$ and $\pm_{\check{\beta}_l}$ unless one of them is known a priori.

\subsubsection{Method to lift sign ambiguity}
\label{subsec_sign_method}

With this background, we can conceive the following procedure to lift all ambiguities in our original problem of estimating $\mathbf{Z}$ for an $N$-port system with $N_\mathrm{S}$ NDA ports.

For the first additional measurement, we connect the $m$th accessible port and the $i$th NDA port to a 2PLN. We terminate all other NDA ports with arbitrary known individual loads. Since there is no sign ambiguity about the mutual impedances between the $m$th accessible port and the other accessible ports, and the measurable impedance matrix is insensitive to $\pm_{\beta_q} \ \forall \ q \neq i$, this setup allows us to determine $\pm_{\beta_i}$. Specifically, we predict the measurable impedance matrix for the two possible values of $\pm_{\beta_i}$ and retain the one yielding the prediction that is closer to the corresponding measurement.

For the second additional measurement, we connect the $i$th and $j$th NDA ports to a 2PLN. We terminate all other NDA ports with arbitrary known individual loads. Since the measurable impedance matrix is insensitive to $\pm_{\beta_q} \ \forall \ q \notin \{i, j\}$ and we know $\pm_{\beta_i}$ from the previous step, this setup allows us to determine $\pm_{\beta_j}$. Specifically, we predict the measurable impedance matrix for the two possible values of $\pm_{\beta_j}$ and retain the one yielding the prediction that is closer to the corresponding measurement.

For the remaining $N_\mathrm{S}-2$ additional measurements, we proceed similarly. We connect a 2PLN to one NDA port whose sign ambiguity we have previously lifted and one whose sign ambiguity is yet to be determined.

The proposed method requires $N_\mathrm{A} \geq 2$ because in the first step one accessible port is connected to the 2PLN and at least one accessible port must remain free for measurements. 
Without the first additional measurement involving an accessible and an NDA port, we could only align all sign ambiguities such that $\pm_{\beta_1} = \pm_{\beta_2} = \pm_{\beta_i} = \dots$ but we could not know their value. Nonetheless, this would imply $\pm_{\kappa_{ij}} = 1 \ \forall \ \{ i,j \}$ such that we would be able to determine $\mathbf{Z}_\mathcal{SS}$ free of ambiguities. Meanwhile, all entries of $\mathbf{Z}_\mathcal{AS}$ would be subject to the same sign ambiguity. Such a result may be sufficient in applications where only the block $\mathcal{SS}$ is of interest. Then, $N_\mathrm{A} = 1 $ is sufficient.

Altogether, the proposed principled method is simple (it requires only a coaxial cable as 2PLN) and has an acceptable linear complexity scaling (it requires $N_\mathrm{S}$ additional measurements).\footnote{If one was to alternatively directly apply the gradient-descent method from Ref.~\cite{del2024minimal} to a generic tunable load network (such as the one envisioned for a fully-connected BD-RIS), one would get an estimate of $\mathbf{Z}$ (or $\mathbf{S}$) only featuring the same sign ambiguity on all entries of $\mathbf{Z}_\mathcal{AS}$ (or $\mathbf{S}_\mathcal{AS}$). However, knowing the  $N_\mathrm{S}\times N_\mathrm{S}$ impedance matrix for each configuration of a generic tunable $N_\mathrm{S}$-port load network might require an $N_\mathrm{S}$-port VNA in which case it could be prohibitively costly for large $N_\mathrm{S}$.}

\subsection{Blockwise Phase Ambiguity Removal}
\label{sec_phase_ambiguity_removal}

In this section, we consider the case of phase-insensitive intensity-only measurements and work with scattering parameters, in line with the corresponding Sec.~V in Ref.~\cite{del2024minimal}.

\subsubsection{Background on phase ambiguity}
\label{sub_sec_background_phase_ambiguity}

Analogous to the developments in terms of impedance parameters in Sec.~\ref{sub_sec_background_sign_ambiguity}, we first consider a generic $M$-port system terminated by a 2PLN. We can show using
\begin{equation}
\check{\mathbf{S}}^\mathrm{2PLN} = \begin{bmatrix}    e & f \\ g &  h     \\ \end{bmatrix} \in\mathbb{C}^{2 \times 2},
\end{equation}
and the convenient partition
\begin{equation}
    \check{\mathbf{S}} = \begin{bmatrix}  \check{\mathbf{S}}_\mathcal{FF}  &  \check{\mathbf{s}}_k &  \check{\mathbf{s}}_l \\ \check{\mathbf{s}}_k^T & \check{\sigma}_k & \check{\nu}_{kl}   \\ \check{\mathbf{s}}_l^T  & \check{\nu}_{kl} & \check{\sigma}_l   \\ \end{bmatrix}
    \label{eq7}
\end{equation}
that (assuming the 2PLN is reciprocal, i.e., $f=g$)
\begin{equation}
\begin{aligned}
   \hat{\check{\mathbf{S}}} = {\check{\mathbf{S}}}_\mathcal{FF} +  \check{\xi} \left[ (h - \check{\sigma}_l)\check{\mathbf{s}}_k\check{\mathbf{s}}_k^T + (e-\check{\sigma}_k)\check{\mathbf{s}}_l\check{\mathbf{s}}_l^T \right. \\
   \left. - (f-\check{\nu}_{kl}) \left(\check{\mathbf{s}}_k\check{\mathbf{s}}_l^T + \check{\mathbf{s}}_l\check{\mathbf{s}}_k^T \right) \right],
\end{aligned}
\label{eq8}
\end{equation}
where 
\begin{equation}
    \check{\xi} = \left((e-\check{\sigma}_k)(h-\check{\sigma}_l) - (f-\check{\nu}_{kl})^2\right)^{-1}.
\end{equation}
Similarly to Sec.~\ref{sub_sec_background_sign_ambiguity}, there is a sign ambiguity about $\check{\mathbf{s}}_k$ [$\check{\mathbf{s}}_l$] which we denote by $\pm_{\check{\gamma}_k}$ [$\pm_{\check{\gamma}_l}$] and about $\check{\nu}_{kl}$ which we denote by $\pm_{\check{\nu}_{kl}}$. Furthermore, $\pm_{\check{\nu}_{kl}}=\pm_{\check{\gamma}_k} \pm_{\check{\gamma}_l}$.

We now return to our original problem of determining $\mathbf{S}$ for an $N$-port system with $N_\mathrm{S}$ NDA ports. 
A restriction to intensity-only measurements results in an additional blockwise phase ambiguity originating from the fact that the magnitudes of $\mathrm{e}^{\jmath \theta}\hat{\mathbf{S}}$ are insensitive to $\theta$: 
\begin{equation}
\begin{split}
\left| \hat{\mathbf{S}} \right| = &\left| \mathrm{e}^{\jmath \theta}\mathbf{S}_\mathcal{AA} + \left(\mathrm{e}^{\jmath \theta/2} \mathbf{S}_\mathcal{AS}\right) \left( \mathbf{S}_\mathrm{L}^{-1} - \mathbf{S}_\mathcal{SS} \right)^{-1} \left( \mathrm{e}^{\jmath \theta/2} \mathbf{S}_\mathcal{SA}\right) \right|, \\ \ &\forall \ \theta.
\end{split}
\end{equation}
Specifically, with individually tunable loads on the NDA ports, we can retrieve via gradient descent $\mathrm{e}^{\jmath \theta}\mathbf{S}_\mathcal{AA}$ instead of $\mathbf{S}_\mathcal{AA}$, $\mathrm{e}^{\jmath \theta/2} \mathbf{S}_\mathcal{AS}^\pm$ (where $^\pm$ indicates column-wise sign ambiguities due to $\pm_{\gamma_i}$) instead of $\mathbf{S}_\mathcal{AS}$, and  $\mathbf{S}_\mathcal{SS}^\pm$ (where $^\pm$ indicates symmetric off-diagonal sign ambiguities due to $\pm_{\nu_{ij}}$) instead of $\mathbf{S}_\mathcal{SS}$, where $\theta$ is an unknown phase~\cite{del2024minimal}.

In sight of this blockwise phase ambiguity, let us reconsider the generic problem from the outset of this subsection involving an $M$-port system terminated by a 2PLN for the specific case in which one of the two ports (the one indexed $k$) connected to the 2PLN is an accessible port and the other one (the one indexed $l$) is an NDA port. Due to the blockwise phase ambiguity, our estimate $\check{\mathbf{S}}^\mathrm{PRED}$ of $\check{\mathbf{S}}$ obtained with intensity-only measurements is 
\begin{equation}
    \check{\mathbf{S}}^\mathrm{PRED} = \begin{bmatrix}  \mathrm{e}^{\jmath \theta}\check{\mathbf{S}}_\mathcal{FF}  &  \mathrm{e}^{\jmath \theta}\check{\mathbf{s}}_k &  \mathrm{e}^{\jmath \theta/2}\check{\mathbf{s}}_l \\ \mathrm{e}^{\jmath \theta}\check{\mathbf{s}}_k^T & \mathrm{e}^{\jmath \theta}\check{\sigma}_k & \mathrm{e}^{\jmath \theta/2}\check{\nu}_{kl}   \\ \mathrm{e}^{\jmath \theta/2}\check{\mathbf{s}}_l^T  & \mathrm{e}^{\jmath \theta/2}\check{\nu}_{kl} & \check{\sigma}_l   \\ \end{bmatrix}
    \label{eq77}
\end{equation}
which implies
\begin{equation}
    \check{\xi}^\mathrm{PRED} = \left((e-\mathrm{e}^{\jmath \theta}\check{\sigma}_k)(h-\check{\sigma}_l) - (f-\mathrm{e}^{\jmath \theta/2}\check{\nu}_{kl})^2\right)^{-1}
    \label{eq777}
\end{equation}
and
\begin{equation}
\begin{split}
   \hat{\check{\mathbf{S}}}^\mathrm{PRED} = \mathrm{e}^{\jmath \theta} \Big[ {\check{\mathbf{S}}}_\mathcal{FF} +  \check{\xi}^\mathrm{PRED} \Big( \mathrm{e}^{\jmath \theta} (h - \check{\sigma}_l)
   \check{\mathbf{s}}_k\check{\mathbf{s}}_k^T  \\
   + (e-\mathrm{e}^{\jmath \theta}\check{\sigma}_k)
   \check{\mathbf{s}}_l\check{\mathbf{s}}_l^T 
   - \mathrm{e}^{\jmath \theta/2}(f-\mathrm{e}^{\jmath \theta/2}\check{\nu}_{kl}) 
   \left(\check{\mathbf{s}}_k\check{\mathbf{s}}_l^T + \check{\mathbf{s}}_l\check{\mathbf{s}}_k^T \right) 
   \Big) \Big].
\end{split}
\label{eq888}
\end{equation}
Various terms appearing in Eq.~(\ref{eq777}) and Eq.~(\ref{eq888}) are sensitive to the exact value of $\theta$, such that in general $\left|  \hat{\check{\mathbf{S}}}^\mathrm{PRED} \right|$ equals $ \left|\hat{\check{\mathbf{S}}}\right|$ only for $\theta=0$.

\subsubsection{Method to lift phase ambiguity}
\label{subsec_phase_method}

Using the same $N_\mathrm{S}$ additional measurement setups involving a 2PLN as in Sec.~\ref{sec_sign_ambiguity} but using intensity-only measurements here, we can lift both the sign and blockwise phase ambiguities. However, we proceed in the opposite order. First, we align all sign ambiguities such that all entries of $\mathbf{S}_\mathcal{AS}$ and $\mathbf{S}_\mathcal{SA}$ have the same sign ambiguity and that $\mathbf{S}_\mathcal{SS}$ is free of ambiguities. Then, we determine $\theta$ and fix the blockwise phase ambiguity that is specific to working with intensity-only data.

For the first additional measurement, we connect the first and second NDA ports to the 2PLN. We terminate the other NDA ports by arbitrary known individual loads. Then, we measure the intensities of the measurable scattering matrix. The predictions of the latter are insensitive to the value of $\theta$ as well as to $\pm_{\check{\gamma}_i} \ \forall \ i>2$. We predict the measurable scattering matrix intensities for the two possible values of $\pm_{\check{\gamma}_2}$, and retain the one yielding the prediction that is closer to the corresponding measurement. Thereby, we impose $\pm_{\check{\gamma}_2} = \pm_{\check{\gamma}_1}$. 

For the next $N_\mathrm{S}-2$ measurements, we proceed similarly. We connect the 2PLN to one NDA port to which it has been connected previously and to one NDA port to which it has not been connected previously, in order to decide whether to flip the sign of the column and row corresponding to the latter. 

Finally, we connect the 2PLN to the $m$th accessible port and the first NDA port. We terminate the remaining NDA ports with known arbitrary individual loads and measure the intensities of the measurable scattering matrix. We sweep the predictions of the latter through all possible values of $\theta$ and retain the value of $\theta$ for which the difference between the intensities of measured and predicted scattering coefficients are simultaneously approaching zero for each scattering coefficient. We observed numerically that examining a single scattering coefficient may yield various possible values of $\theta$; therefore, it is important to ensure $N_\mathrm{A}>2$ in the case of intensity-only measurements such that the measurable scattering matrix is at least of dimensions $2 \times 2$ in the last measurement, providing three distinct scattering coefficients.

\begin{figure*}[h]
    \centering
    \includegraphics[width=0.9\linewidth]{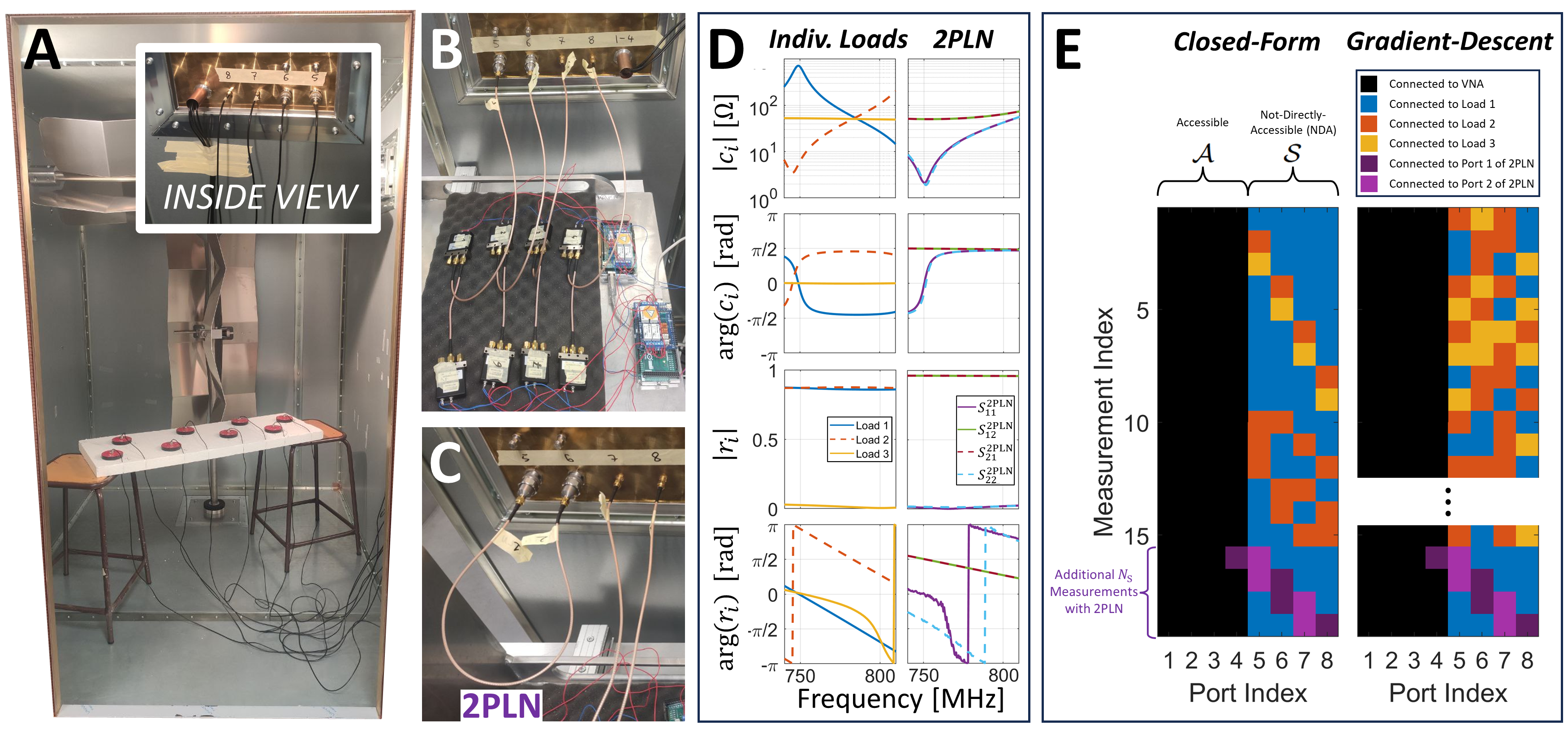}
    \caption{Experimental setup. (A) Reverberation chamber comprising 8 antennas. (B) Setup to electronically switch the termination of four ports between three individual loads. (C) Setup involving the 2PLN (a coaxial cable). (D) Impedance and scattering characteristics of the three individual loads (left column) and the four coefficients characterizing the 2PLN (right column). (E) Measurement procedure for the ``Virtual VNA 2.0''. $\mathcal{A}$ and $\mathcal{S}$ identify the accessible and NDA ports, respectively. The schematic shows whether a given port is connected to the VNA (black) or terminated by one of the three individual loads (blue, red, yellow) or the 2PLN (purple). The closed-form approach (left column) requires a series of specific configurations, the gradient-descent approach (right column) can be applied to an arbitrarily long sequence of random configurations (except for the last few involving the 2PLN).}
    \label{Fig1}
\end{figure*}

\section{Experimental Results}
\label{sec_exp}

For our proof-of-principle experiments, our DUT consists of an array of eight antennas (AEACBK081014-S698) with monomodal SubMiniature version A (SMA) ports placed inside a reverberation chamber, as shown in Fig.~\ref{Fig1}A. Thereby, it is clear that we cannot use any a priori knowledge about the DUT (e.g., about symmetries or other geometric details). Moreover,  the chamber's through-wall SMA connectors enable a perfectly stable setup on which the terminations of the DUT ports can be changed without perturbing the DUT. Furthermore, we can easily measure the ground truth with our eight-port VNA (two cascaded Keysight P5024B four-port VNAs) to validate our results. These are hence ideal conditions for the presented proof-of-principle experiment in which we split the eight DUT ports into four accessible ports and four NDA ports. As shown in Fig.~\ref{Fig1}B, we use electro-mechanical relay switches (PE71S6436) to reconfigure the NDA ports' individual loads. None of the three realizable individual loads emulates a perfect calibration standard (matched load, open circuit, or short circuit) across the considered frequency interval from 740~MHz to 810~MHz, as seen in Fig.~\ref{Fig1}D. To realize connections between pairs of DUT ports via a 2PLN, we manually connect them to a simple coaxial cable, as seen in Fig.~\ref{Fig1}C. For each considered load configuration (illustrated in Fig.~\ref{Fig1}E for both the closed-form and the gradient-descent approaches), we measure the $4\times 4$ scattering matrix at the DUT's four accessible ports with our VNA (IFBW:~500~kHz; power:~13~dBm). (For the configuration involving the connection of one accessible and one NDA port via the 2PLN, of course, we only measure a $3 \times 3$ scattering matrix.)

Because none of the realizable individual loads emulates a perfect calibration standard, the closed-form approach requires an additional data analysis step that was proposed and numerically validated in Sec.~III-B of Ref.~\cite{del2024minimal}; our subsequent results constitute the first experimental validation of this additional processing step.

We quantify the accuracy of our estimated impedance or scattering matrix with a metric quantifying the ratio between signal and estimation error:
\begin{equation}
    \zeta = \left\langle\frac{\mathrm{SD}\left[Z_{ij}^\mathrm{GT}(f)\right]}{\mathrm{SD}\left[Z_{ij}^\mathrm{GT}(f) - Z_{ij}^\mathrm{PRED}(f)\right]}\right\rangle_{i,j,f},
    \label{eq_zeta}
\end{equation}
where $\mathrm{SD}$ denotes the standard deviation, and the superscripts GT and PRED denote ground truth and prediction, respectively. When we estimate scattering rather than impedance parameters, we replace $Z_{ij}$ by $S_{ij}$ in Eq.~(\ref{eq_zeta}).

\subsection{Sign Ambiguity Removal}

\begin{figure}[b]
    \centering
    \includegraphics[width=0.75\linewidth]{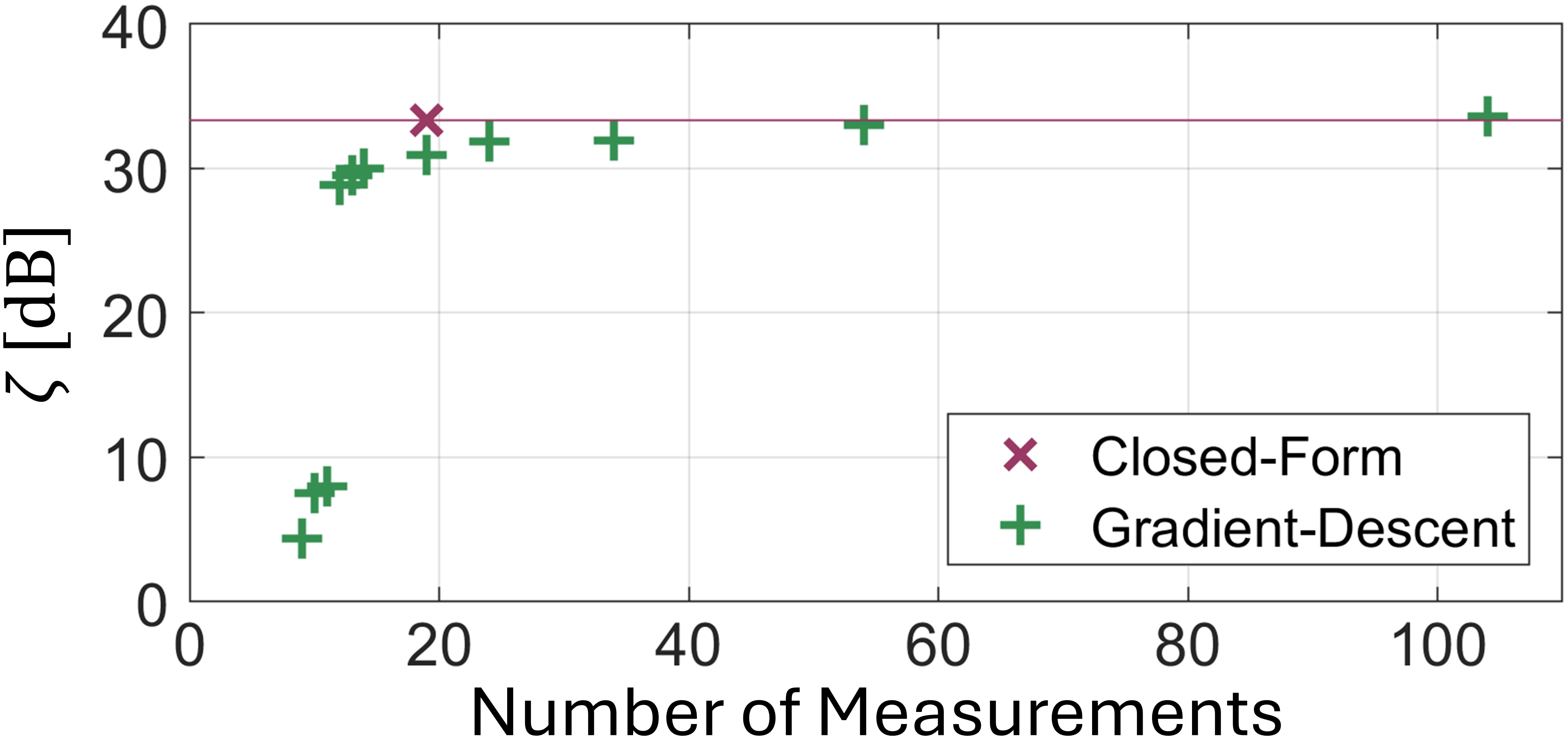}
    \caption{Experimentally achieved frequency-averaged accuracy $\zeta$ (see Eq.~(\ref{eq_zeta}) for definition) with complex-valued measurements (averaged over 141 linearly spaced frequency points between 740~MHz and 810~MHz).}
    \label{Fig2}
\end{figure}

The experimentally achieved accuracies with the closed-form and gradient-descent approaches based on complex-valued measurements at the accessible ports are displayed in Fig.~\ref{Fig2} as a function of the number of utilized measurements. In contrast to Ref.~\cite{del2024minimal}, we observe that the closed-form approach is very competitive in terms of achieved accuracy. We attribute this observation to two facts. On the one hand, the reproducibility of the electronic switching used here is much better than that of the manual switching used in Ref.~\cite{del2024minimal} for the closed-form approach. 
On the other hand, the results are obtained in the low-noise regime. The gradient-descent approach can be expected to become more advantageous as the level of noise increases because it can flexibly adapt the number of measurements and because the changes of the measurable scattering matrix are more significant (since more than one or two NDA port terminations are changed between any given pair of measurements)~\cite{del2024minimal}.

Importantly, despite the lower reproducibility of manual switching in comparison to electronic switching, our reliance on manual switching for the $N_\mathrm{S}$ additional measurements involving the 2PLN did not pose a problem because these additional measurements are only used for binary sign decisions (as elaborated in Sec.~\ref{subsec_sign_method}). Hence, the required accuracy for these additional measurements involving the 2PLN is lower than for the other measurements.

\subsection{Blockwise Phase Ambiguity Removal}

As mentioned, we use the same experiment but assume to be restricted to phaseless measurements now. We follow the procedure based on the gradient-descent method for intensity-only measurements detailed in Ref.~\cite{del2024minimal} except for an improved procedure that aligns the phase offsets (see Appendix~F in Ref.~\cite{del2024minimal}): we take the median of $\theta^\mathrm{B}$ obtained for multiple (rather than only one) previously aligned realizations.

\begin{figure}[b!]
    \centering
    \includegraphics[width=\columnwidth]{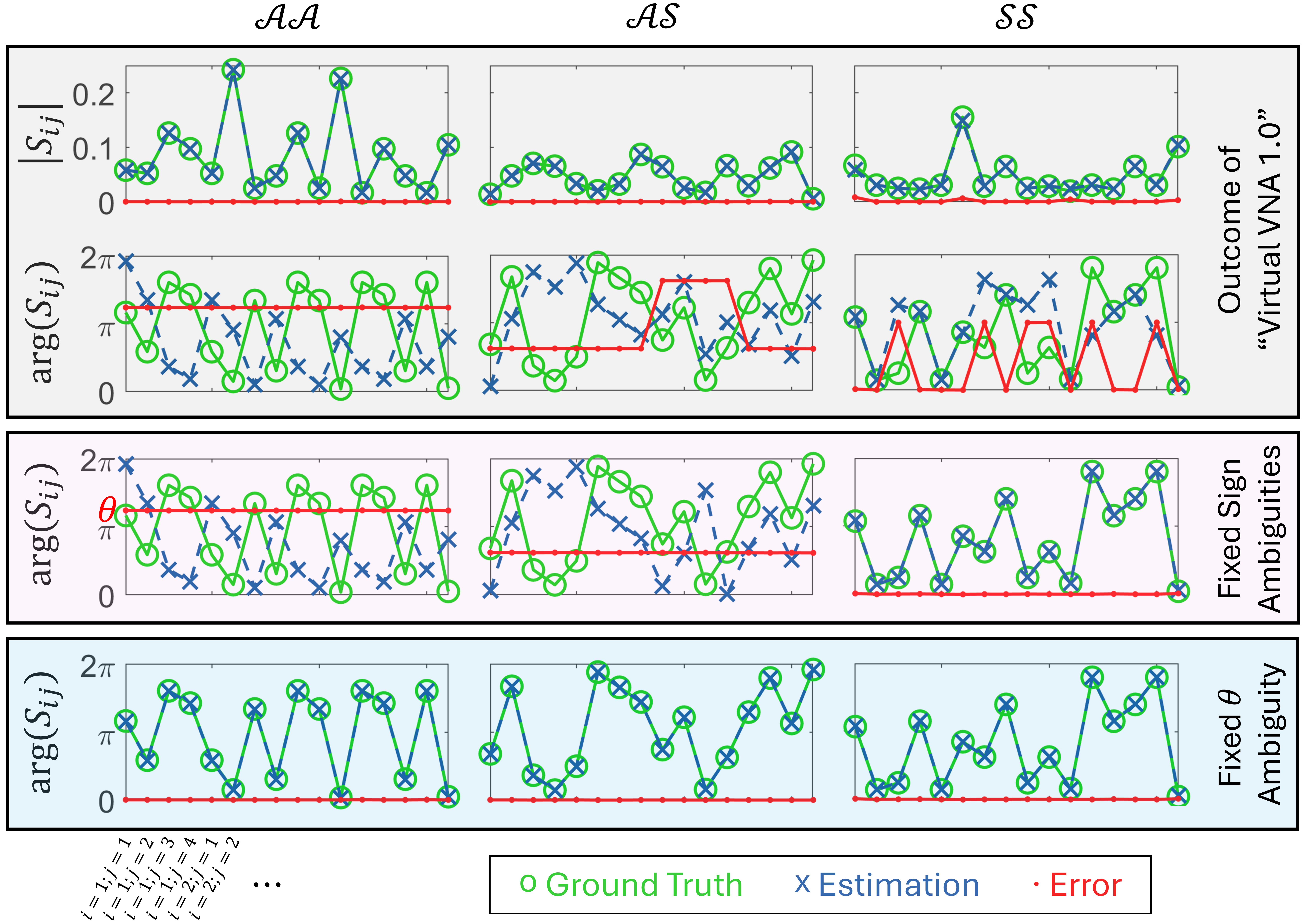}
    \caption{Step-by-step procedure for ambiguity-free estimation of the full scattering matrix purely based on intensity-only measurements for 771~MHz. The three columns display $\mathbf{S}_\mathcal{AA}$, $\mathbf{S}_\mathcal{AS}$ and $\mathbf{S}_\mathcal{SS}$ in vectorized form. }
    \label{FigPL}
\end{figure}

Results of the various steps of the procedure outlined in Sec.~\ref{subsec_phase_method} are displayed in Fig.~\ref{FigPL} for the 771~MHz frequency point. The combined sign and blockwise phase ambiguities after the first step are apparent, and the remaining blockwise phase ambiguity after the second step is apparent, too. The final outcome is free of ambiguities and reaches an accuracy of 32~dB.

Unlike the sign alignment, the estimation of $\theta$ to remove the blockwise phase ambiguity is not only a binary decision and hence significantly more vulnerable to reproducibility issues. Indeed, we encountered such issues at some frequency points. Future work will overcome these issues with a fully electronic switching setup.

\vspace{-1.5mm}

\section{Discussion}
\label{sec_discussion}

With complex-valued measurements, the ``Virtual VNA 2.0'' requires $N_\mathrm{A}>1$ and the number of required configurations depends on the chosen approach. The closed-form approach (suitable in the low-noise regime) fixes an upper bound of $1+3N_\mathrm{S}+N_\mathrm{S}(N_\mathrm{S}-1)/2$ on the required number of configurations under ideal (low-noise) conditions. With intensity-only measurements, we found that the ``Virtual VNA 2.0'' requires $N_\mathrm{A}>2$ and we are not aware of an upper bound on the required number of configurations.

The main limitation of the ``Virtual VNA 2.0'' concept is its vulnerability to noise because it inherently relies on measuring minute changes of the measurable scattering matrix as a function of load configuration.\footnote{A systematic noise sensitivity analysis can be found in Fig.~4A of Ref.~\cite{del2024minimal}.} One mitigation of this limitation consists in using more accessible ports. Moreover, in case of noisy measurements, it is advisable to use the gradient-descent approach because the changes of the measurable scattering matrix therein are typically larger and because additional measurements for new random sets of load configurations can be incorporated smoothly. Meanwhile, imprecisions inherent to the manual manipulation of the loads are easily eliminated with a fully electronic switching setup.

\vspace{-1.5mm}

\section{Conclusion}
\label{sec_conclusion}

We have theoretically and experimentally demonstrated the ``Virtual VNA 2.0'' which can estimate the full scattering matrix of an arbitrarily complex and unknown system without ever connecting a subset of $N_\mathrm{S}$ system ports to the VNA. The previously proposed Virtual VNA~\cite{del2024minimal} solely used individual tunable loads to terminate the NDA ports such that it could not remove vexing sign ambiguities, and an additional blockwise phase ambiguity in the case of phase-insensitive measurements. Here, we worked out and experimentally validated that MPLNs are the key to lifting these ambiguities. We found a very simple implementation, requiring only $N_\mathrm{S}$ additional measurements with a simple coaxial cable acting as 2PLN (i.e., the additional measurement complexity scales linearly with $N_\mathrm{S}$).

Looking forward, we envision that the two embodiments of the ``Virtual VNA 2.0'' concept illustrated in Fig.~\ref{Fig_sketch}B,C will be realized with fully automated load switching to tackle the challenging characterization of large and/or embedded antenna arrays; a very
recent PCB prototype matching the ``Virtual VNA 2.0'' embodiment in Fig.~\ref{Fig_sketch}B can be found in~\cite{KitVNA}.

\vspace{-2mm}

\bibliographystyle{IEEEtran}
\providecommand{\noopsort}[1]{}\providecommand{\singleletter}[1]{#1}%

\end{document}